\documentclass{article}
\usepackage[utf8]{inputenc}
\usepackage{parskip}
\usepackage{amsmath}
\usepackage{amssymb}
\usepackage{setspace}
\PassOptionsToPackage{normalem}{ulem}
\usepackage{ulem}
\makeatletter
\@ifundefined{date}{}{\date{}}
\title{Effective Degrees of Freedom for Balanced Repeated Replication and Jackknife Variance Estimates: A Unified Approach via Stratum Contrasts}
\author{}

\makeatother

\begin{document}
\title{Effective Degrees of Freedom for Balanced Repeated Replication and
Paired Jackknife Variance Estimates: A Unified Approach via Stratum
Contrasts}
\author{Matthias von Davier}
\maketitle
\begin{abstract}
Balanced repeated replication (BRR) and the jackknife are two widely
used methods for estimating variances in stratified samples with two
primary sampling units per stratum. While both methods produce variance
estimators that can be expressed as sums of squared stratum-level
contrasts, they differ fundamentally in their construction and in
the dependence structure of their replicate estimates. This article
examines the independence properties of the components contributing
to these variance estimators. For BRR, we show that although the replicate
estimates themselves are correlated, the balancing property of Hadamard
matrices collapses the variance estimator into a sum of independent
stratum-specific components. For the jackknife, the independence of
components follows directly from the construction. Using these independence
results, we derive the variance of each variance estimator and establish
a direct connection to the Welch--Satterthwaite degrees of freedom
approximation. This yields a practical formula for estimating degrees
of freedom when constructing confidence intervals for population totals.
The derivation highlights the unified treatment of both replication
methods and provides insights into their relative efficiency and applicability. 
\end{abstract}

\section{Introduction}

In complex sample surveys, variance estimation is essential for constructing
confidence intervals and conducting hypothesis tests. For stratified
designs where each stratum contains exactly two primary sampling units
(PSUs), replication methods offer a flexible and design-consistent
approach to variance estimation. Two prominent methods in this context
are balanced repeated replication (BRR) and the jackknife repeated
replication (JRR). Despite their different constructions, both methods
produce variance estimators that can be expressed in terms of within-stratum
contrasts, yet they exhibit distinct dependence structures among their
replicate estimates.

BRR, introduced by McCarthy (1966), creates replicates by systematically
selecting one PSU from each stratum according to the rows of a Hadamard
matrix. This balanced selection ensures that each stratum contributes
to every replicate, and the orthogonality of the Hadamard columns
simplifies the resulting variance estimator. However, because the
replicates share data across all strata, the replicate estimates are
correlated, raising questions about the effective degrees of freedom
for inference.

The jackknife, in contrast, creates replicates by deleting one PSU
at a time and adjusting weights accordingly. In the two-PSU-per-stratum
design, this yields $2H$ replicate estimates, each modifying only
a single stratum. While the replicates within a stratum are perfectly
correlated, the contributions from different strata are independent
by construction.

A remarkable fact, often underappreciated, is that both the BRR variance
estimator $\hat{V}_{\text{BRR}}=\frac{1}{R}\sum_{r}\left(\hat{T}_{r}-\hat{T}\right)^{2}$
and the jackknife variance estimator $\hat{V}_{\text{JRR}}=\sum_{h}\frac{1}{2}\left[\left(\hat{T}_{(h1)}-\hat{T}\right)^{2}+\left(\hat{T}_{(h2)}-\hat{T}\right)^{2}\right]$
reduce algebraically to the same simple expression: $\sum^{H}_{h=1}d^{2}_{h}$,
where $d_{h}=w_{h1}y_{h1}-w_{h2}y_{h2}$ is the within-stratum contrast.
This common representation reveals that both estimators are sums of
independent stratum-level components, despite the differing dependence
structures of their underlying replicates.

This article makes three contributions. First, we derive the covariance
structure of replicate deviations for BRR, explicitly showing how
the balancing property eliminates cross-stratum dependencies in the
variance estimator. Second, we analyze the variance of the variance
estimator itself, expressing it in terms of fourth moments of the
stratum contrasts. Third, and most importantly, we connect these results
to the Welch--Satterthwaite (W-S) degrees of freedom approximation.
By treating each $d^{2}_{h}$ as an independent component with approximately
one degree of freedom, we obtain a practical formula for the effective
degrees of freedom: 
\[
\hat{\nu}=3\frac{\bigl(\sum^{H}_{h=1}d^{2}_{h}\bigr)^{2}}{\sum^{H}_{h=1}d^{4}_{h}}-2.
\]
This formula is based on von Davier's (2026) work on a bias correction
for the W-S equation, and applies equally to BRR and the jackknife,
providing a unified approach to degrees of freedom estimation for
both replication methods.

The remainder of the paper is organized as follows. Section 2 describes
an exemplary sampling design used in BRR and introduces notation.
Section 3 presents the BRR construction and derives the covariance
structure of replicate deviations. Section 4 does the same for the
jackknife. Section 5 examines the variance of the squared deviations
and establishes the connection to the W-S equation. Section 6 discusses
practical implications and concludes.

\section{Sampling Design and Notation}

Consider a finite population partitioned into $H$ strata. Within
each stratum $h=1,\dots,H$, two primary sampling units (PSUs) are
selected, possibly with unequal probabilities. Let $w_{hi}$ denote
the sampling weight for unit $i$ ($i=1,2$) in stratum $h$, and
let $y_{hi}$ be the estimate of the variable of interest. The full-sample
estimator of the population total is 
\[
\hat{T}=\sum^{H}_{h=1}\sum^{2}_{i=1}w_{hi}y_{hi}.
\]

Define the within-stratum contrast 
\[
d_{h}=w_{h1}y_{h1}-w_{h2}y_{h2},
\]
which captures the difference between the two weighted observations
in stratum $h$. Under the stratified design, the units drawn in each
of the strata are independent, so the random variables $d_{h}$ are
independent across $h$. We assume that the sampling design is such
that $E[d_{h}]=0$ for each stratum, a condition that holds under
simple random sampling within strata and more generally when the sampled
units are representative of the stratum population. The variance of
$\hat{T}$ is then 
\[
V\left(\hat{T}\right)=\sum^{H}_{h=1}\text{Var}\left(d_{h}\right).
\]

\section{Balanced Repeated Replication}

\subsection{Construction of BRR Replicates}

Balanced repeated replication uses a Hadamard matrix to create replicates.
A Hadamard matrix $\mathbf{H}$ of order $R$ (where $R$ is a multiple
of 4) is an $R\times R$ matrix with entries $\alpha_{rh}\in\{-1,+1\}$
satisfying $\mathbf{H}'\mathbf{H}=R\mathbf{I}$. For a design with
$H$ strata, we select the first $H$ columns of a Hadamard matrix
of order $R\ge H$.

For replicate $r=1,\dots,R$, the entry $\alpha_{rh}$ determines
which unit is selected in stratum $h$: if $\alpha_{rh}=+1$, unit
1 receives doubled weight $2w_{h1}$ and unit 2 receives weight zero;
if $\alpha_{rh}=-1$, unit 2 receives doubled weight $2w_{h2}$ and
unit 1 receives weight zero.

The replicate estimate of the total is 
\[
\hat{T}_{r}=\sum^{H}_{h=1}\left(\frac{1+\alpha_{rh}}{2}\cdot2w_{h1}y_{h1}+\frac{1-\alpha_{rh}}{2}\cdot2w_{h2}y_{h2}\right)
\]

\[
=\sum^{H}_{h=1}\left(\left[w_{h1}y_{h1}+w_{h2}y_{h2}\right]+\alpha_{rh}\left[w_{h1}y_{h1}-w_{h2}y_{h2}\right]\right)
\]

\[
=\hat{T}+\sum^{H}_{h=1}\alpha_{rh}d_{h}
\]

note that, since $\frac{1\pm\alpha_{rh}}{2}\in\left\{ 0,1\right\} $,
we have $2\frac{1\pm\alpha_{rh}}{2}\in\left\{ 0,2\right\} .$ Also
note that the mean of the replicates equals the estimate of the total,
that is
\[
\frac{1}{R}\sum^{R}_{r=1}\hat{T}_{r}=\hat{T}+\frac{1}{R}\sum^{H}_{h=1}d_{h}\sum^{R}_{r=1}\alpha_{rh}=\hat{T}
\]
since the $R$ are even and by definition $\sum^{R}_{r=1}a_{rh}=0.$
Then, we define the replicate deviate as $X_{r}=\left(\hat{T}_{r}-\hat{T}\right)=\sum^{H}_{h=1}\alpha_{rh}d_{h}$.

\subsection{Covariance Structure of BRR Deviations}

Because the strata are independent and $E[d_{h}]=0$, the covariance
between two replicate deviations is 
\[
\text{Cov}(X_{r},X_{s})=\sum^{H}_{h=1}\alpha_{rh}\alpha_{sh}\text{Var}(d_{h}).
\]

The rows of a Hadamard matrix are orthogonal: $\sum^{H}_{h=1}\alpha_{rh}\alpha_{sh}=0$
for $r\neq s$. However, this does not imply that $\text{Cov}(X_{r},X_{s})=0$
unless all $\text{Var}(d_{h})$ are equal. In general, the replicate
deviations are correlated, with the covariance depending on the alignment
of the sign patterns with the stratum variances.

\subsection{Sum of Squared Deviations}

The squared deviation for replicate $r$ is 
\[
X^{2}_{r}=\left(\sum^{H}_{h=1}\alpha_{rh}d_{h}\right)^{2}=\sum^{H}_{h=1}\sum^{H}_{k=1}\alpha_{rh}\alpha_{rk}d_{h}d_{k}.
\]

Summing over all replicates and using the column orthogonality property
$\sum^{R}_{r=1}\alpha_{rh}\alpha_{rk}=R$ if $h=k$ and $0$ otherwise,
we obtain 

\[
\sum^{R}_{r=1}X^{2}_{r}=\sum^{R}_{r=1}\sum^{H}_{h=1}\sum^{H}_{k=1}\alpha_{rh}\alpha_{rk}d_{h}d_{k}=\sum^{H}_{h=1}\sum^{H}_{k=1}d_{h}d_{k}\sum^{R}_{r=1}\alpha_{rh}\alpha_{rk}
\]
 and then 
\[
\sum^{H}_{h=1}\sum^{H}_{k=1}d_{h}d_{k}\sum^{R}_{r=1}\alpha_{rh}\alpha_{rk}=\sum^{H}_{h=1}d_{h}d_{h}\left(\sum^{R}_{r=1}\alpha_{rh}\alpha_{rh}\right)
\]
 and hence
\[
\sum^{R}_{r=1}X^{2}_{r}=R\sum^{H}_{h=1}d^{2}_{h}.
\]

Thus, the BRR variance estimator equals
\[
\hat{V}_{\text{BRR}}=\frac{1}{R}\sum^{R}_{r=1}\left(\hat{T}_{r}-\hat{T}\right)^{2}_{r}=\sum^{H}_{h=1}d^{2}_{h}
\]
and is a sum of \uline{independent} stratum-level components, the
squared stratum level contrasts, despite the correlation among the
replicates $\hat{T}_{r}$ themselves.

\section{Jackknife Repeated Replication}

\subsection{Construction of Jackknife Replicates}

In the jackknife approach for two PSUs per stratum, we create $2H$
replicates by deleting one unit at a time. For stratum $h$ and unit
$i$, the replicate estimate $\hat{T}_{(hi)}$ is obtained by dropping
unit $i$ from stratum $h$ and doubling the weight of the remaining
unit in that stratum, while keeping all other strata unchanged. The
deviation for this replicate is 
\[
\hat{T}_{(hi)}-\hat{T}=\begin{cases}
-d_{h} & \text{if }i=1\text{ (unit 1 deleted)}\\
+d_{h} & \text{if }i=2\text{ (unit 2 deleted)}.
\end{cases}
\]

This can be seen as follows. Recall that $\hat{T}=\sum^{H}_{h=1}\sum^{2}_{i=1}w_{hi}y_{hi}$
and that 
\[
\hat{T}_{(hi)}=\sum_{j\ne h}\sum^{2}_{l=1}w_{jl}y_{jl}+2I_{\left\{ 1\right\} }\left(i\right)w_{h1}y_{h1}+2I_{\left\{ 2\right\} }\left(i\right)w_{h2}y_{h2}
\]
so that 
\[
\hat{T}_{(hi)}-\hat{T}=2I_{\left\{ 1\right\} }\left(i\right)w_{h1}y_{h1}+2I_{\left\{ 2\right\} }\left(i\right)w_{h2}y_{h2}-w_{h1}y_{h1}-w_{h2}y_{h2}.
\]
If the indicator function $I_{\left\{ 1\right\} }\left(i\right)=1$
then $I_{\left\{ 2\right\} }\left(i\right)=0$ and $\hat{T}_{(hi)}-\hat{T}=2w_{h1}y_{h1}+0w_{h2}y_{h2}-w_{h1}y_{h1}-w_{h2}y_{h2}=d_{h},$and
accordingly $\hat{T}_{(hi)}-\hat{T}=-d_{h}$ if $I_{\left\{ 1\right\} }\left(i\right)=0$
and $I_{\left\{ 2\right\} }\left(i\right)=1$. 

\subsection{Independence Structure}

The quantities $d_{h}$ are independent across strata because each
involves only units from within a single stratum. Within a given stratum
$h$, the two deviations $\hat{T}_{(h1)}-\hat{T}$ and $\hat{T}_{(h2)}-\hat{T}$
are perfectly correlated (they are negatives of each other). However,
this within-stratum dependence does not affect the variance estimator,
which combines information across strata, as will be seen below.

The standard jackknife variance estimator for this 2-PSU per stratum
design is 
\[
\hat{V}_{\text{JRR}}=\sum^{H}_{h=1}\frac{1}{2}\left[\left(\hat{T}_{(h1)}-\hat{T}\right)^{2}+\left(\hat{T}_{(h2)}-\hat{T}\right)^{2}\right]=\sum^{H}_{h=1}d^{2}_{h}.
\]

Thus, the jackknife variance estimator reduces to exactly the same
expression as the BRR estimator. In this case, the independence of
the components $d^{2}_{h}$ follows directly from the construction,
as each $d_{h}$ is based on a distinct stratum and strata are independent.

\section{Fay's Method for Balanced Repeated Replication and the Paired Jackknife}

Fay (1989) introduced a modification to the BRR and jackknife variance
estimators to address issues arising from assigning zero weights in
replicate estimates. In standard BRR and the paired jackknife, each
replicate assigns a weight of zero to one unit in each affected stratum.
This can be problematic for subpopulation analyses, where units with
zero weights are effectively deleted from the replicate, potentially
leading to instability or undefined estimates for small domains. Fay's
method replaces the $\{0,2\}$ weighting scheme with a perturbation
factor $\varepsilon$ (typically $\varepsilon=0.5$), assigning weights
of $(1+\varepsilon)w_{hi}$ and $(1-\varepsilon)w_{hi}$ to the two
units in each stratum. This ensures that all units retain positive
weights in every replicate, while preserving the expectation of the
replicate estimates and the unbiasedness of the variance estimator.

\subsection{BRR with Fay's Method}

In the BRR context, the replicate weight for unit $i$ in stratum
$h$ for replicate $r$ is given by 
\[
w^{(r)}_{hi}=w_{hi}\bigl(1+\alpha_{rh}\delta_{hi}\varepsilon\bigr),
\]
where $\alpha_{rh}\in\{-1,+1\}$ is the Hadamard matrix entry, $\delta_{h1}=+1$,
$\delta_{h2}=-1$, and $\varepsilon$ is the perturbation factor ($0<\varepsilon\le1$). 

As an example, for $\varepsilon=0.5$ we obtain

\[
w^{(r)}_{h1}=w_{h1}\left(1+0.5\alpha_{rh}\right)\in\left\{ 0.5,1.5\right\} \ni w_{h2}\left(1-0.5\alpha_{rh}\right)=w^{(r)}_{h2}.
\]
The replicate estimate of the total then becomes 
\[
\hat{T}_{r}=\sum^{H}_{h=1}\sum^{2}_{i=1}w_{hi}y_{hi}\bigl(1+\alpha_{rh}\delta_{hi}\varepsilon\bigr)=
\]

\[
\hat{T}+\varepsilon\sum^{H}_{h=1}\alpha_{rh}\sum^{2}_{i=1}w_{hi}y_{hi}\delta_{hi}=\hat{T}+\varepsilon\sum^{H}_{h=1}\alpha_{rh}d_{h}.
\]

with $d_{h}=w_{h1}y_{h1}-w_{h2}y_{h2}$ as defined earlier. The deviation
from the full-sample estimate is therefore 
\[
\hat{T}_{r}-\hat{T}=\varepsilon\sum^{H}_{h=1}\alpha_{rh}d_{h}.
\]

The standard BRR variance estimator using Fay's method incorporates
a correction factor $1/\varepsilon^{2}$ to account for the scaling:
\[
\hat{V}_{\text{Fay-BRR}}=\frac{1}{\varepsilon^{2}R}\sum^{R}_{r=1}\left(\hat{T}_{r}-\hat{T}\right)^{2}.
\]
Substituting the expression for the deviations and using the orthogonality
of the Hadamard columns, 
\[
\sum^{R}_{r=1}(\hat{T}_{r}-\hat{T})^{2}=\varepsilon^{2}\sum^{R}_{r=1}\left(\sum^{H}_{h=1}\alpha_{rh}d_{h}\right)^{2}=\varepsilon^{2}R\sum^{H}_{h=1}d^{2}_{h},
\]
so that 
\[
\hat{V}_{\text{Fay-BRR}}=\frac{1}{\varepsilon^{2}R}\cdot\varepsilon^{2}R\sum^{H}_{h=1}d^{2}_{h}=\sum^{H}_{h=1}d^{2}_{h}.
\]
Thus the Fay-BRR variance estimator coincides with the standard BRR
estimator $\hat{V}_{\text{BRR}}=\sum_{h}d^{2}_{h}$, independent of
the choice of the Fay term $0<\varepsilon\le1$. Trivially, for $\varepsilon=1$
we obtain the regular\textbf{ }BRR described above. 

\subsection{Paired Jackknife with Fay's Method}

For the paired jackknife with two units per stratum, Fay's method
creates two replicates per stratum. In replicate $(h1)$, the units
$i=1,2$ receive weights $(1+\delta_{hi}\varepsilon)w_{hi}$; in replicate
$(h2)$, the assignments are reversed with weights $(1-\delta_{hi}\varepsilon)w_{hi}$.
All other strata retain their original weights. The replicate estimates
are 
\[
\hat{T}_{(h1)}=\hat{T}+\varepsilon d_{h},\qquad\hat{T}_{(h2)}=\hat{T}-\varepsilon d_{h}.
\]
The Fay jackknife variance estimator is 
\[
\hat{V}_{\text{Fay-JK}}=\frac{1}{2\varepsilon^{2}}\sum^{H}_{h=1}\left[\left(\hat{T}_{(h1)}-\hat{T}\right)^{2}+\left(\hat{T}_{(h2)}-\hat{T}\right)^{2}\right].
\]
Because each squared deviation equals $\varepsilon^{2}d^{2}_{h}$,
we obtain 
\[
\hat{V}_{\text{Fay-JK}}=\frac{1}{2\varepsilon^{2}}\sum^{H}_{h=1}\left(2\varepsilon^{2}d^{2}_{h}\right)=\sum^{H}_{h=1}d^{2}_{h},
\]
again matching the standard jackknife estimator.

\subsection{Implications for Degrees of Freedom}

The key observation from the derivations above is that Fay's method
does not alter the fundamental expression of the variance estimator
as $\hat{V}=\sum^{H}_{h=1}d^{2}_{h}$. The $d_{h}$ are independent
across strata, and their squares constitute independent components
of the variance estimate. Consequently, the W-S degrees of freedom
approximation applies unchanged to Fay's method. The corrected equation
for the estimated degrees of freedom, 
\[
\hat{\nu}=\frac{3\bigl(\sum^{H}_{h=1}d^{2}_{h}\bigr)^{2}}{\sum^{H}_{h=1}d^{4}_{h}}-2,
\]
remains valid for constructing confidence intervals and conducting
hypothesis tests based on $\hat{V}_{\text{Fay-BRR}}$ or $\hat{V}_{\text{Fay-JK}}$.

Fay's method offers practical advantages beyond the theoretical equivalence.
By avoiding zero weights, it ensures that every replicate can be used
to estimate totals for any subpopulation, even those with small sample
sizes. The perturbation factor $\varepsilon$ can be tuned to balance
the trade-off between stability and the magnitude of the replicate
deviations; $\varepsilon=0.5$ is a common choice, but other values
(e.g., $\varepsilon=0.3$ or $0.7$) may be employed depending on
the design and analysis goals. Importantly, the unbiasedness of the
variance estimator is preserved for any $\varepsilon\neq0$, as shown
above.

In summary, Fay's method provides a flexible and robust extension
of BRR and the paired jackknife that retains the essential independence
structure of the stratum contrasts and therefore integrates seamlessly
with the degrees of freedom framework developed in this article.

\section{Variance of the Variance Estimator and Degrees of Freedom}

\subsection{Variance of the Squared Deviations}

The derivations above have shown that we have $\sum^{R}_{r=1}X^{2}_{r}=R\sum^{H}_{h=1}d^{2}_{h}$.
The variance of this sum over repeated replicated samples is 
\[
\text{Var}\left(\sum^{R}_{r=1}X^{2}_{r}\right)=\text{Var}\left(R\sum^{H}_{h=1}d^{2}_{h}\right)=R^{2}\sum^{H}_{h=1}\text{Var}\left(d^{2}_{h}\right)
\]

Under the assumption $E[d_{h}]=0$, $\text{Var}\left(d^{2}_{h}\right)=E\left[d^{4}_{h}\right]-\left(E\left[d^{2}_{h}\right]\right)^{2}$.
If, in addition, we assume that within each stratum the observations
are approximately normal or that the sample sizes are large enough
for central limit theorem effects, we may approximate $d^{2}_{h}\sim\sigma^{2}_{h}\chi^{2}_{1}$,
where $\sigma^{2}_{h}=E\left[d^{2}_{h}\right]$. Then $\text{Var}\left(d^{2}_{h}\right)=2\sigma^{4}_{h}$.

\subsection{Connection to the W-S Equation}

Both replication methods yield the variance estimator $\hat{V}=\sum^{H}_{h=1}d^{2}_{h}$,
a sum of independent random variables. To construct confidence intervals
for $\hat{T}$, we often assume that $\hat{T}$ is approximately normal
and that $\hat{V}$ is approximately a scaled chi-squared random variable:
$\frac{\nu\hat{V}}{\sigma^{2}}\sim\chi^{2}_{\nu}$ with $\sigma^{2}=E\left(\hat{V}\right)$
and $\nu$ degrees of freedom. 

An estimate of $\nu$ is obtained by the W-S equation (Satterthwaite,
1941, 1946; Welch, 1947): 
\[
\nu=\frac{\left(\sum^{H}_{h=1}\sigma^{2}_{h}\right)^{2}}{\sum^{H}_{h=1}\sigma^{4}_{h}}.
\]

In practice, the unknown $\sigma^{2}_{h}$ are replaced by their estimates
$d^{2}_{h}$, giving the estimated degrees of freedom 
\[
\hat{\nu}=\frac{\left(\sum^{H}_{h=1}d^{2}_{h}\right)^{2}}{\sum^{H}_{h=1}d^{4}_{h}},
\]
or, alternatively, by applying the bias-corrected version
\[
\hat{\nu}=\frac{3\left(\sum^{H}_{h=1}d^{2}_{h}\right)^{2}}{\sum^{H}_{h=1}d^{4}_{h}}-2
\]

as derived by von Davier (2026). This formula applies directly to
both BRR and the jackknife, providing a unified method for degrees
of freedom estimation. A confidence interval for the population total
can then be constructed as 
\[
\hat{T}\pm t_{\hat{\nu},1-\alpha/2}\sqrt{\hat{V}},
\]
where $t_{\hat{\nu},1-\alpha/2}$ is the appropriate quantile from
the $t$-distribution with $\hat{\nu}$ degrees of freedom.

\section{Discussion and Practical Implications}

The equivalence of the BRR and jackknife variance estimators to the
simple sum $\sum_{h}d^{2}_{h}$ has important practical implications.
First, it demonstrates that despite their different constructions,
both methods capture the same essential information about stratum-level
variability. Second, the independence of the $H$ components $d^{2}_{h}$
justifies the use of the W-S approximation for degrees of freedom,
which accounts for the heterogeneity of variances across strata. Third,
the Satterthwaite equation cannot directly be applied to the $2H$
replicates $T_{\left(hi\right)}-T$ used in the paired jackknife,
since this would involve using each of the $d^{2}_{h}=\left(T_{h1}-T\right)^{2}=\left(T_{h2}-T\right)^{2}$twice,
hence introducing correlated variance components. Instead, it is required
to use
\[
\hat{\nu}=\frac{3\left(\sum^{H}_{h=1}\left(T_{\left(h1\right)}-T\right)^{2}\right)}{\sum^{H}_{h=1}\left(T_{\left(h1\right)}-T\right)^{4}}-2
\]

Fourth, the application of Fay's method does not lead to a different
expression for the estimator needed for the determination of the effective
degrees of freedom when applying the W-S equation. 

Fifth, in case of the simple (JK1; delete-a-group) jackknife, with
one unit per zone, and dropping each jackknife zone once, the deviates
$\delta_{(i)}=\left(\hat{T}_{i}-\hat{T}\right)^{2}$can be viewed
as (approximately, depending on whether and how weights are adjusted
for each replicate) independent, each $\chi^{2}$distributed with
with approximately one degree of freedom, and the (corrected) W-S
equation directly applied to $S^{2}=\sum^{H}_{i=1}\delta^{2}_{(i)}$. 

Sixth, unfortunately, a similarly simple solution cannot be directly
applied to the BRR replicate deviations $X_{r}=\left(\hat{T}_{r}-\hat{T}\right)=\sum^{H}_{h=1}\alpha_{rh}d_{h}$
since these are linear combinations of the within stratum contrasts
$d_{h}=w_{h1}y_{h1}-w_{h2}y_{h2}.$ However, given that the weights
$w_{hi}$ are known for each of the units, these stratum contrasts
are easily computed, and the W-S equation, or its bias corrected alternative,
can be directly applied to the components of $S^{2}_{BRR}=\sum_{h}d^{2}_{h}.$

In applications, the estimated degrees of freedom $\hat{\nu}$ will
typically be less than $H$ when the stratum variances are unequal,
and can be as low as 1 in extreme cases. This reflects the loss of
information due to variance heterogeneity and provides a more accurate
reflection of the uncertainty in the variance estimate than simply
using $H$ degrees of freedom.

For BRR, it is noteworthy that the balancing property of the Hadamard
matrix effectively "decorrelates" the replicate contributions when
aggregated into the variance estimator, even though the individual
replicates are correlated. This insight clarifies why BRR, despite
its complex replicate structure, can be treated analogously to the
jackknife for degrees of freedom purposes.

\section*{References}

Fay, R. E. (1989). Theory and application of replicate weighting for
variance calculations. In \textit{Proceedings of the Survey Research
Methods Section}, American Statistical Association, 212--217.

McCarthy, P. J. (1966). Replication: An approach to the analysis of
data from complex surveys. \textit{Vital and Health Statistics}, Series
2, No. 14. National Center for Health Statistics.

Satterthwaite, F. E. (1946). An approximate distribution of estimates
of variance components. \textit{Biometrics Bulletin}, 2(6), 110--114.

Rao, J. N. K., and Shao, J. (1992). Jackknife variance estimation
with survey data under hot deck imputation. \textit{Biometrika}, 79(4),
811--822.

Tukey, J. W. (1958). Bias and confidence in not-quite large samples.
\textit{Annals of Mathematical Statistics}, 29(2), 614--623.

von Davier, M. (2026). A corrected Welch Satterthwaite equation. And:
What you always wanted to know about Kish's effective sample but were
afraid to ask. arXiv:2602.20912

Welch, B. L. (1947). The generalization of 'Student's' problem when
several different population variances are involved. \textit{Biometrika},
34(1/2), 28--35.Wolter, K. M. (2007). \textit{Introduction to Variance
Estimation} (2nd ed.). Springer.
\end{document}